# Global research trends and collaborations in Fibrodysplasia Ossificans Progressiva: A bibliometric analysis (1989-2023)

Muneer Ahmad, *PhD*[1], Undie Felicia Nkatv, *PhD*[2] and Sajid Saleem[3]
[1]Chief University Librarian, [2]Ag. Deputy Librarian, [3]Lecturer,
[1,2]The Iddi Basajjabalaba Memorial Library, Kampala International University, Box 20000, Ggaba Road, Kansanga, Uganda
[3]Department IT, SOMAC, Kampala International University, Box 20000, Ggaba Road, Kansanga, Uganda
E-mail: [1]muneerbangroo@gmail.com; ahmad.muneer@kiu.ac.ug; [2]nkavfelicia@gmail.com; felicia.undie@kiu.ac.ug; [3]sajid.saleem@kiu.ac.ug.
Mobile: [1]+256709415657; [2]+2348025668229
Orcid: [1]https://orcid.org/0000-0001-5666-8438; [2]https://orcid.org/0009-0002-0496-8505; [3]https://orcid.org/0009-0002-6131-8576

**Abstract**
Fibrodysplasia Ossificans Progressiva (FOP) is a rare and debilitating genetic disorder characterized by the progressive formation of bone in muscles and connective tissues. This scientometric analysis examines the global research trends on FOP between 1989 and 2023 using bibliographic data from Web of Science. The study highlights key patterns in publication productivity, influential journals, institutions, and the geographical distribution of research. The findings reveal that the United States leads both in terms of total publications and citation impact, with significant contributions from the UK, Italy, Japan, and other European countries. Additionally, the analysis identifies the major document types, including articles and reviews, and evaluates the collaborative efforts across institutions. The study offers valuable insights into the global research landscape of FOP, providing a foundation for future studies and international collaborations.

## Introduction

Fibrodysplasia ossificans progressive (FOP), which is also referred to as myositis ossificans (Nilsson, 1998) represents a rare autosomal dominant genetic disorder characterized by an incidence rate of approximately one in two million live births, exhibiting no discernible sexual, racial, or geographical predisposition (Miao et al., 2012). The majority of affected individuals are dispersed globally, with the exception of certain cases involving familial inheritance (Pignolo et al., 2016). The initial accounts of FOP, documented by Patin (1692) and Freke (1739), delineate its clinical manifestations (Shore, Gannon, & Kaplan, 1997). Subsequently, Stonham, Burton-Fanning, and various other medical practitioners have documented cases involving patients of diverse genders, age groups, and even entire familial units exhibiting the phenotypic characteristics associated with FOP (Shore et al., 2006).

Unusual formation of bone in the joints and soft tissues, such as muscles, tendons, and ligaments, are common signs of FOP (Hildebrand et al., 2017). This condition can lead to disabilities like skeletal deformities, chronic pain, and restricted mobility, greatly impacting the patients' quality of life and mental well-being. The life expectancy for FOP patients is generally no more than 40 years (Zhang et al., 2013). The specific pathogenesis of FOP is not yet clear, and the early phenotype of the disease is easily confused with other diseases, including tumors, fibromas, and bursitis, resulting in its misdiagnosis (Kitterman et al., 2005). Diagnosis of FOP can be challenging due to its similarity to other conditions, and there are currently no effective treatments available (Kaplan et al., 2018).

FOP is a very uncommon genetic disease that is autosomal dominant and affects 1 out of every 2,000,000 people. Ninety-five percent of patients experience heterotopic ossification (HO) before the age of 15, but there has been a report of a patient with HO at 56 years old [(Whyte et al., 2012); Baujat et al., 2017). Data from the CEMARA and PMSI databases shows that the average age of FOP patients is 25.5 years, the average age of onset is 7.1 years, and the average age at diagnosis is 10.2 years (Baujat et al., 2017).

Data on Europe shows that the UK has confirmed 30 cases out of 49 million residents, giving a prevalence of 0.61 per million. Spain has an estimated incidence of 0.36 per million (Morales-Piga et al., 2012), and France has a prevalence of 1.36 per million. These numbers are comparable to the international prevalence of the condition (Baujat et al., 2017).

Currently, the majority of reported patients with FOP are in the United States, making up approximately 25.6% of all registered patients. Following closely behind is China, with around 10.8% of registered patients, and Brazil with about 8.4% of patients. Compared to patients in Europe and the Americas, Asian patients tend to be younger (Pignolo et al., 2016). Despite FOP being extremely rare, China still has a significant number of patients due to its large population. While exact figures for China are not known, the prevalence of FOP can be used to estimate the number of patients.

According to reports, there are approximately 650 individuals in China who have been diagnosed with FOP (Zhang et al., 2013). However, only around 70 cases have been officially reported due to limited medical research on the condition. The reported cases make up no more than 12% of the total number of FOP patients in the country. There is a need for improved understanding of the symptoms and mutations associated with FOP, as well as better diagnostic measures for the condition.

**Literature review**

Examining the available literature is an essential part of comprehending the present state of research, recognizing areas where knowledge is lacking, and setting the stage for any new study. The systematic analysis of scientific output is crucial in mapping and assessing the scholarly output of a specific field. This approach not only showcases trends, important contributors, and influential publications but also gives a deeper understanding of how the field has evolved and the patterns of collaboration within it. By consolidating this information, scientometric studies make a substantial contribution to advancing research and guiding future investigations.

Numerous studies have explored the pathomechanisms of depressive disorders and antidepressants, yet scientometric analyses on this topic remain limited. This review employs scientometric methods and a historical overview to examine research trends on depression between 1998 and 2018. The analysis focuses on identifying key keywords, research hotspots, and emerging trends. Findings reveal the field's distribution, knowledge structure, and the evolution of research topics, highlighting themes such as induction factors, comorbidity, pathogenesis, therapy, and animal models. These elements collectively enhance understanding of the occurrence, progression, and treatment of depressive disorders. The study identified 231,270 publications on depression, reflecting a fourfold growth over two decades, providing a comprehensive framework for future research in the field (Xu et al., 2021).

In recent years, migraine has garnered significant attention from researchers worldwide. This study employed scientometric methods to analyze research trends and advancements in migraine studies from 2010 to 2019 using the Web of Science core collection database. Tools such as VOSviewer, CiteSpace, and Excel facilitated the analysis, identifying 6,357 publications (5,203 articles and 1,154 reviews). The United States led in publications (n=2,151, 33.84%), with Albert Einstein College of Medicine contributing the most (n=220, 3.46%). Cephalalgia emerged as the leading journal, with 766 publications and the highest co-citations (n=35,535). Notable contributors included Lipton RB with the most publications (n=159, 2.50%) and Silberstein SD with the most co-citations (n=4,215). Key research areas encompassed causes, pathophysiology, epidemiology, diagnostic criteria, treatments, prevention strategies, and genetic studies related to migraine. This scientometric review offers a comprehensive overview of the field, serving as a valuable resource for future investigations (Lu et al., 2021).

Bipolar disorder (BD) is a significant psychiatric condition with rising global prevalence. While extensive preclinical and clinical studies have explored its pathological mechanisms and pharmacological treatments, a scientometric analysis of research trends and interdisciplinary developments in this field has been lacking. This study conducted a scientometric review of 55,358 publications on BD from 2002 to 2021 to identify key keywords, research hotspots, and trends. The analysis highlighted the field's distribution, knowledge structure, topic evolution, and emerging themes. Key focus areas included risk factors, pathogenesis, critical brain regions, comorbidities, and treatment approaches, offering valuable insights into BD's aetiology, progression, and management. These findings provide a robust framework for advancing research in bipolar disorder (Zhu et al., 2023).

Tardive dyskinesia (TD), recognized as a significant drug-induced condition since the 1960s, has been extensively studied in terms of its clinical features, epidemiology, pathophysiology, and treatment. This scientometric review analyzed 5,228 TD-related publications and 182,052 citations from the Web of Science through 2021 to identify research trends and hotspots. Using VOSviewer and CiteSpace, the study mapped annual output, prominent authors, institutions, and countries, revealing peaks in TD research during the 1990s, a decline post-2004, and renewed interest after 2015. Key contributors included Kane JM, Lieberman JA, and Jeste DV historically, with Zhang XY and Correll CU leading recent studies. The Journal of Clinical Psychiatry and Journal of Psychopharmacology were the most prolific outlets. Early clusters explored TD's clinical and pharmacological characteristics, with later research focusing on oxidative stress, pharmacogenetics, and treatment advances like vesicular monoamine transporter-2 inhibitors since 2017. This review provides a comprehensive resource for understanding TD's research evolution and identifying future directions (Baminiwatta & Correll, 2023).

In the last decade, significant advancements have been made in using artificial intelligence (AI) to address biomedical challenges, particularly in oncology. With over 6,000 publications in this domain, a bibliometric analysis is essential to understand its structure, growth, and trends. This study, the first of its kind, analyzed 6,757 documents on AI in oncology published between 2012 and 2022 using data from the Web of Science. China led in output (2,087 publications, 30.89%), with Sun Yat-sen University as the top institution and Wei Wang as the most prolific author. Scientific Reports published the most articles, while PLOS ONE had the highest co-citations. Emerging research themes included tissue microarrays, segmentation, and artificial neural networks, reflecting the prominence of deep learning in areas like radiomics, genomics, risk stratification, and therapy response. These findings map global collaborations at individual, institutional, and national levels, providing a roadmap for future AI applications in precision oncology and other cutting-edge areas (Wu et al., 2022).

This study examines the status and trends in pressure ulcer care research over the past two decades through bibliometric and visual analysis, aiming to inform clinical practices and future investigations. A search of the Web of Science Core Collection database from 2002 to 2021 yielded 5,102 publications, including 4,034 articles across 1,557 journals, with an annual growth rate of 4.89%. Analysis revealed four main research clusters: epidemiology, risk assessment, care interventions, and prognostic analysis. The "Braden score" emerged as a prominent keyword, while recent research highlights include knowledge dissemination, training, and pressure ulcer care for COVID-19 patients. The findings suggest an increasing focus on education and specialized care, with future research likely to address pressure ulcers in the elderly population due to global aging trend (Li et al., 2024).

This study presents a scientometric and visual analysis of meningioma research from 2012 to 2021, highlighting trends and providing insights for clinical and research applications. Using data from the Web of Science Core Collection, 10,397 documents were analyzed, with 6,714 articles showing an annual growth rate of 8.9%. The United States led in publication output (1,671 articles), with the University of California, San Francisco, as the top institution (242 articles). Research clusters were categorized into five main areas: meningioma characteristics, surgical treatment, radiation therapy, stereotactic radiosurgery, and complication management. Recent trends point to emerging hotspots in meningioma classification and molecular characteristics. The findings underline a decade-long focus on surgical and radiation treatments, with increasing attention to molecular studies, signaling future research directions in meningioma management (Zhang, Feng, Liu, & Liu, 2023).

This study analyzes esketamine research trends from 2000 to 2020 through bibliometric methods. Data from 683 publications were gathered from the Web of Science Core Collection, focusing on annual publication trends, citation metrics, contributing countries, institutions, authors, journals, and keyword patterns. The United States led with the highest number of publications (162, 23.72%), citations (36.08%), and an H-index of 40, while Chiba University in Japan emerged as the top institution. Publications were mainly featured in Anesthesia & Analgesia. Keyword analysis revealed a shift in research focus from anesthesia and analgesia to depression treatment, with depression-related keywords being most prominent. The findings highlight a significant increase in esketamine research since 2016, with emerging research hotspots centering on its efficacy in treating depression. The growth in contributions from Asian countries also underscores their increasing impact on the field (Li, Xiang, Liang, Deng, & Du, 2022).

This bibliometric analysis examined 3,243 publications on male osteoporosis from 1998 to 2020, aiming to map its knowledge framework, identify research hotspots, and predict emerging trends. Male osteoporosis research saw an initial increase in publications, followed by a decrease in recent years. The USA made the most significant contributions, with the highest number of publications and the greatest H-index value. Oregon Health and Science University emerged as the leading institution. Osteoporosis International was the most influential journal in this field. Keywords were grouped into four clusters: basic research, epidemiology and risk factors, diagnostic studies, and treatment and fracture prevention. Burst detection highlighted terms such as "obesity," "zoledronic acid," "DXA," "inflammation," "fall," "microarchitecture," and "sarcopenia" as future research priorities. The study underscores the importance of addressing male osteoporosis as a growing global health concern, especially with the aging population. Continued exploration of risk factors, innovative diagnostics, and advanced treatments is crucial to improving outcomes. This analysis serves as a foundational reference for researchers and policymakers to better understand trends and inform future research directions in male osteoporosis (Wu, Sun, Tong, Wang, Yan, & Sun, 2021).

This bibliometric analysis of global research on non-motor symptoms (NMS) in Parkinson's disease (PD) from 2013 to 2022 revealed a significant increase in publications, with 3,521 articles identified. The United States led in publications and citations, while King's College London was the most active institution. Key research hotspots included early diagnosis, biomarkers, novel MRI techniques, and deep brain stimulation. Alpha-synuclein emerged as the largest research cluster, with non-motor symptoms, quality of life, dementia, and depression being the most frequently studied topics. This analysis highlights trends in PD research, emphasizing the need for advanced diagnostic tools and interventions to address NMS and improve patient care (Li, Chen, Pan, Zhou, Sun, Zhang, et al., 2024).

This bibliometric analysis examines the use of biological disease-modifying anti-rheumatic drugs (bDMARDs) for treating axial spondyloarthritis (axSpA) from 2004 to 2022. A total of 2,546 articles were included, with the USA leading in publications, followed by Germany and the Netherlands. The most frequent publisher was Rheumazentrum Ruhrgebiet, and Annals of the Rheumatic Diseases had the highest number of publications. Research hotspots included tuberculosis, IL-17, and quality of life until 2020, with newer trends focusing on biosimilars, JAK inhibitors, spinal radiographic progression, adverse events, and machine learning. The number of studies has steadily increased, reflecting ongoing advancements in axSpA treatment (He et al., 2023).

There is a lack of comprehensive scientometric evaluation of the research on Fibrodysplasia Ossificans Progressiva (FOP), a rare genetic disorder. Previous literature review shows no prior scientometric studies focused on FOP, indicating a critical gap in understanding research trends, funding patterns, and collaborative efforts in this area. A dedicated scientometric study is necessary to gain valuable insights and identify opportunities for advancing research on FOP.

**Objectives of the study**
The objectives of this study are to:
1. analyze the distribution of publications on Fibrodysplasia Ossificans Progressiva (FOP) from 1989 to 2023;
2. identify the leading journals publishing research on FOP and their citation metrics;
3. examine the most relevant institutions and countries contributing to FOP research;
4. assess the types of documents most frequently published on FOP; and

5. evaluate the collaborative efforts across research institutions and countries in advancing FOP studies.

**Methods**

This scientometric study analyzes research on Fibrodysplasia Ossificans Progressiva (FOP) using bibliographic data retrieved from the Web of Science database. The search was conducted using the terms "Myositis Ossificans" OR "Fibrodysplasia Ossificans Progressiva," which yielded a total of 1,817 records. The Web of Science, a highly reputable and comprehensive database, was chosen for its extensive coverage of academic publications and citation data.

To analyze the retrieved data, multiple tools were utilized: Histcite was employed to generate citation-based analyses, Bibexcel was used for data manipulation and analysis, and MS Excel for organizing and visualizing the data. VOSviewer was used to create visual representations of bibliometric networks, including citation relationships and institutional collaborations. This combination of tools allowed for a detailed analysis of the publication trends, collaborative efforts, and citation metrics related to FOP research from 1989 to 2023.

**Results and discussion**

**Table 1: Annual distribution of publications and citations on Fibrodysplasia Ossificans Progressiva**

| S.No. | Year | Records | % | Rank | TLCS | % | TGCS | % |
|---|---|---|---|---|---|---|---|---|
| 1 | 1989-1993 | 59 | 3.25 | 7 | 967 | 7.28 | 1700 | 3.78 |
| 2 | 1994-1998 | 143 | 7.87 | 6 | 1716 | 12.92 | 4081 | 9.06 |
| 3 | 1999-2003 | 151 | 8.31 | 5 | 1338 | 10.08 | 5209 | 11.57 |
| 4 | 2004-2008 | 191 | 10.51 | 4 | 2552 | 19.22 | 9140 | 20.30 |
| 5 | 2009-2013 | 294 | 16.18 | 3 | 3372 | 25.39 | 9365 | 20.80 |
| 6 | 2014-2018 | 411 | 22.62 | 2 | 2678 | 20.17 | 12079 | 26.82 |
| 7 | 2019-2024 | 568 | 31.26 | 1 | 656 | 4.94 | 3458 | 7.68 |
| | Total | 1817 | 100.00 | | 13279 | 100.00 | 45032 | 100.00 |

Table 1 provides a detailed account of the annual distribution of publications and citations on Fibrodysplasia Ossificans Progressiva (FOP) from 1989 to 2024. It outlines the trends in publication volumes and their corresponding citation impact, represented by Total Local Citation Score (TLCS) and Total Global Citation Score (TGCS). Over the 36-year span, 1,817 publications were recorded, with a progressive increase in scholarly output, reflecting growing interest and awareness of the condition within the scientific community.

From 1989 to 1993, a total of 59 papers were created, making up 3.25% of the overall output. These papers received 967 TLCS and 1,700 TGCS, indicating a moderate impact both within the dataset and on a global scale. Despite having the lowest number of publications, this phase of research laid the groundwork for future studies. In the following period, from 1994 to 1998, the number of publications almost tripled to 143, accounting for 7.87% of the total output. This increase was accompanied by a significant rise in TLCS (1,716 or 12.92%) and TGCS (4,081 or 9.06%), showing a growing recognition of FOP research within and beyond academic circles. Between 1999 and 2003, the number of publications experienced a small increase to 151 (8.31%). Despite the slight growth in volume, the TLCS declined to 1,338 (10.08%), while the TGCS had a significant rise to 5,209 (11.57%). This period seems to indicate a phase where the influence of earlier studies continued to grow globally, even as newer publications had a somewhat reduced local citation impact. The subsequent period, from 2004 to 2008, represented a significant milestone in FOP research, with publications rising to 191 (10.51%). This time frame also witnessed the highest TLCS (2,552 or 19.22%), along with a substantial surge in TGCS to 9,140 (20.30%). These measures suggest that crucial research findings were released during this period, resulting in substantial scholarly attention.

From 2009 to 2013, there was a notable increase in the number of publications, with 294 records (16.18%), indicating a surge in research activity. During this time, there were also the highest TLCS (3,372 or 25.39%) and one of the highest TGCS (9,365 or 20.80%), highlighting its significance within the scientific community. It is likely that this period saw important studies that advanced the understanding of FOP and influenced future research directions. Between 2014 and 2018, the number of publications peaked at 411 records (22.62%), and while TLCS decreased slightly to 2,678 (20.17%), TGCS reached its highest value of 12,079 (26.82%). This indicates that research during this time had a significant global impact, possibly due to wider dissemination and adoption of findings by the international research community.

From 2019 to 2024, there has been a significant rise in the number of published works, with 568 records showing a 31.26% increase. On the other hand, TLCS has decreased to 656 (4.94%), and TGCS has dropped to 3,458 (7.68%). This decrease in citation metrics is probably due to the shorter period for these publications to accumulate citations. It is anticipated that these figures will increase as the newer studies become more widely recognized over time.

In general, the data shows a steady increase in the number of publications, revealing a continuous and expanding interest in FOP research. The citation metrics point out important time periods, especially from 2009 to 2018, when the field made significant progress. The recent decrease in citation scores underscores the importance of monitoring the lasting effects of recent research. This examination offers valuable perspectives on the development of FOP research, pinpointing significant periods of advancement and influence that have molded the field.

Table 2 provides information about the publishing and citation impact of the top 20 authors in Fibrodysplasia Ossificans Progressiva (FOP) research. Collectively, these authors have produced 1,223 publications, representing 67.31% of the total dataset. Their Total Local Citation Score (TLCS) is 16,848, and their Total Global Citation Score (TGCS) is 32,420. These figures highlight the significant role of these authors in advancing FOP research, demonstrated by their extensive publication output and substantial citation metrics.

Dr. Kaplan FS is recognized as the most prolific author, with 226 published papers, representing 12.44% of the total publications. His work has garnered the highest TLCS (5,022) and TGCS (9,539), demonstrating his significant influence in shaping the field and the global reach of his research. In a close second is Dr. Shore EM, who has authored 152 publications (8.37%), with a TLCS of 3,656 and TGCS of 7,657. The considerable citation impact highlights her substantial contributions to advancing the understanding of FOP at both local and international levels. Dr. Pignolo RJ takes the third position with 101 publications (5.56%), achieving a TLCS of 1,152 and TGCS of 2,280, cementing his importance in the research community.

Table 2: Publication output of top 20 authors and citation score

| S.No. | Authors | Records | % | TLCS | TGCS |
|---|---|---|---|---|---|
| 1 | Kaplan FS | 226 | 12.44 | 5022 | 9539 |
| 2 | Shore EM | 152 | 8.37 | 3656 | 7657 |
| 3 | Pignolo RJ | 101 | 5.56 | 1152 | 2280 |
| 4 | Ranganath LR | 92 | 5.06 | 862 | 1400 |
| 5 | Gallagher JA | 83 | 4.57 | 981 | 1472 |
| 6 | Santucci A | 64 | 3.52 | 860 | 1422 |
| 7 | Hsiao EC | 56 | 3.08 | 163 | 674 |
| 8 | Al Mukaddam M | 48 | 2.64 | 128 | 281 |
| 9 | Keen R | 44 | 2.42 | 98 | 267 |
| 10 | Braconi D | 40 | 2.20 | 697 | 1067 |
| 11 | Bernardini G | 38 | 2.09 | 567 | 913 |
| 12 | Katagiri T | 36 | 1.98 | 672 | 1776 |
| 13 | Baujat G | 35 | 1.93 | 75 | 298 |
| 14 | Milan AM | 34 | 1.87 | 349 | 569 |
| 15 | Millucci L | 34 | 1.87 | 524 | 766 |
| 16 | Eekhoff EMW | 30 | 1.65 | 67 | 448 |
| 17 | Davison AS | 28 | 1.54 | 256 | 441 |
| 18 | Taylor AM | 28 | 1.54 | 318 | 438 |
| 19 | De Cunto C | 27 | 1.49 | 68 | 168 |
| 20 | Hughes AT | 27 | 1.49 | 333 | 544 |
| **Total** | | **1223** | **67.31** | **16848** | **32420** |

The works of distinguished authors such as Dr. Ranganath LR (92 publications), Dr. Gallagher JA (83 publications), and Dr. Santucci A (64 publications) have made a significant impact, evidenced by their TLCS values surpassing 800 and TGCS exceeding 1,400. Their consistent productivity and high citation scores demonstrate enduring influence in the field. On the other hand, Dr. Hsiao EC (56 publications) and Dr. Al Mukaddam M (48 publications), though having fewer publications, have still made an impression, with TGCS values of 674 and 281 respectively, indicating their research is gaining recognition, albeit on a smaller scale.

It is interesting to note that certain authors, like Dr. Katagiri T (36 publications) and Dr. Braconi D (40 publications), have a notably high citation impact (TGCS of 1,776 and 1,067, respectively) in comparison to the amount of work they have produced. This suggests that their research is highly esteemed and frequently referenced within the academic community. Conversely, authors like Dr. Baujat G and Dr. De Cunto C have lower TLCS and TGCS values, which implies a more limited or developing influence in their field.

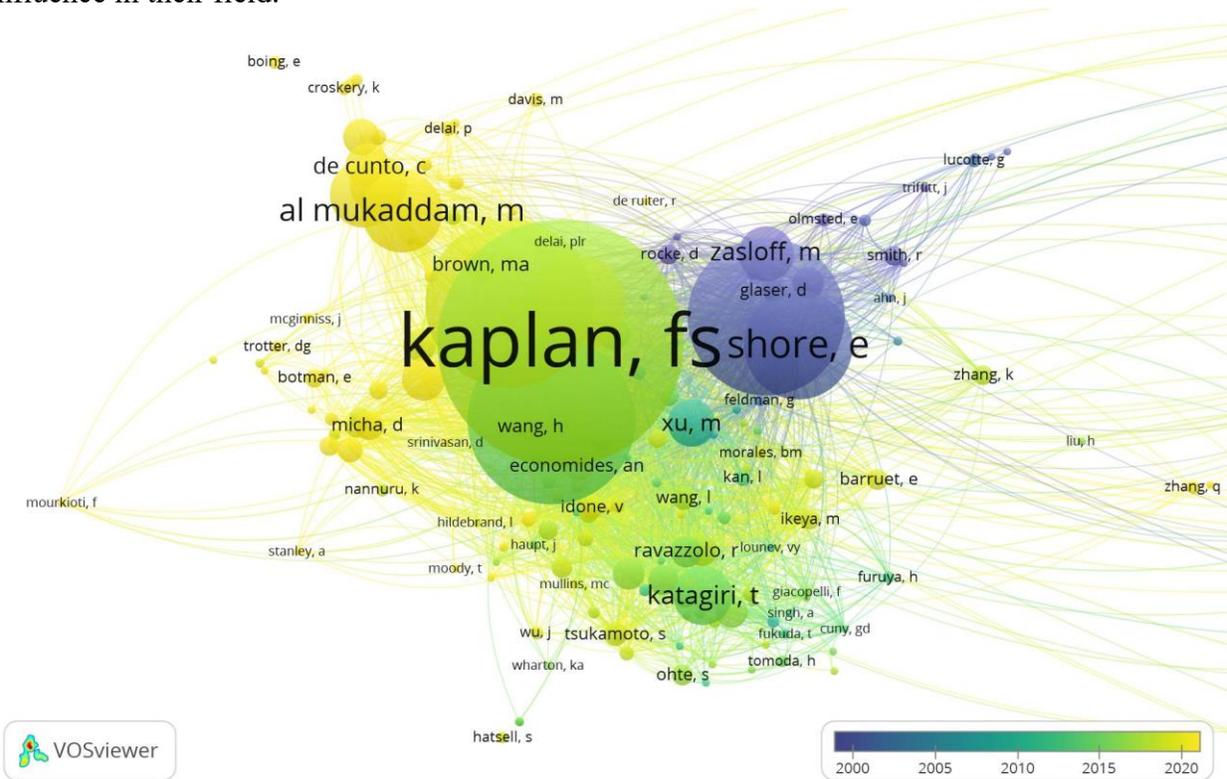

**Figure 1:** Author collaboration network on FOP research

The Total Link Strength (TLS) indicates in figure 1 the extent of collaborative influence among authors. Kaplan FS (14,830) and Shore EM (12,250) dominate as key contributors to collaboration networks. Other authors, such as Pignolo RJ (9,315) and Santucci A (10,084), also demonstrate substantial connectivity, while lower TLS values for Hsiao EC (4,165) and Al Mukaddam M (2,371) suggest less extensive collaborations. The combined work of these writers emphasizes the collaborative aspect of FOP research, showing that a small group of key individuals are responsible for a large portion of scholarly activity and influence. The data further highlights the significant impact of influential authors such as Dr. Kaplan and Dr. Shore, whose extensive research and high citation metrics demonstrate their positions as leaders in the field. However, newer contributors also indicate the constantly changing nature of FOP research, as recent studies contribute to the ongoing development and improvement of scientific knowledge about this uncommon condition.

This examination offers a thorough perspective on the influential contributors and their importance, providing valuable understanding into the cooperative endeavors that have influenced the development and focus of FOP research over time. The diverse citation metrics indicate a mix of foundational research and recent progress, all of which contribute to the continuous expansion of expertise in this specialized field.

Table 3: Top 20 source-wise distribution of publications

| S.No. | Journals | Documents | % | TLCS | TGCS |
|---|---|---|---|---|---|
| 1 | Journal of Bone and Mineral Research | 143 | 7.87 | 808 | 2022 |
| 2 | Bone | 75 | 4.13 | 613 | 1949 |
| 3 | Journal of Inherited Metabolic Disease | 39 | 2.15 | 399 | 665 |
| 4 | Clinical Orthopaedics and Related Research | 33 | 1.82 | 567 | 1035 |
| 5 | Orphanet Journal of Rare Diseases | 26 | 1.43 | 0 | 598 |
| 6 | American Journal of Medical Genetics Part A | 22 | 1.21 | 154 | 500 |
| 7 | Molecular Genetics and Metabolism | 22 | 1.21 | 183 | 410 |
| 8 | Clinical Rheumatology | 21 | 1.16 | 133 | 202 |
| 9 | American Journal of Human Genetics | 19 | 1.05 | 124 | 420 |
| 10 | Osteoarthritis and Cartilage | 18 | 0.99 | 47 | 63 |
| 11 | Rheumatology International | 18 | 0.99 | 91 | 223 |
| 12 | Calcified Tissue International | 17 | 0.94 | 145 | 272 |
| 13 | Journal of Cellular Physiology | 17 | 0.94 | 188 | 401 |
| 14 | Journal of Bone and Joint Surgery-American Volume | 16 | 0.88 | 843 | 2278 |
| 15 | Value in Health | 16 | 0.88 | 4 | 7 |
| 16 | Rheumatology | 15 | 0.83 | 134 | 337 |
| 17 | European Journal of Human Genetics | 14 | 0.77 | 148 | 253 |
| 18 | Annals of the Rheumatic Diseases | 12 | 0.66 | 130 | 177 |
| 19 | Biomedicines | 12 | 0.66 | 0 | 98 |
| 20 | Biomolecules | 11 | 0.61 | 0 | 20 |

Table 3 emphasizes the most important 20 journals that have made a substantial impact on the publication of research on Fibrodysplasia Ossificans Progressiva (FOP). These journals have collectively published 1,817 articles, making them the preferred sources for sharing FOP-related studies. Additionally, these journals vary in their Total Local Citation Scores (TLCS) and Total Global Citation Scores (TGCS), indicating their influence and reach within the scientific community.

The Journal of Bone and Mineral Research is the top source, with 143 publications, accounting for 7.87% of the total contributions. It also has a high citation impact, with a TLCS of 808 and TGCS of 2,022, indicating its importance as a fundamental journal in the field, attracting high-quality and influential research on bone-related disorders, including FOP. Similarly, Bone has published 75 articles (4.13%), with a strong TLCS of 613 and TGCS of 1,949, further highlighting its significance as a primary source for bone and mineral metabolism studies.

Additional prominent publications are the Journal of Inherited Metabolic Disease, which provided 39 articles (2.15%) with a TLCS of 399 and TGCS of 665, and Clinical Orthopaedics and Related Research, which released 33 articles (1.82%) and attained noteworthy citation metrics (TLCS 567, TGCS 1,035). These journals prove their significance in the areas of inherited disorders and clinical orthopedic research, respectively.

The Total Link Strength (TLS) in figure 2 highlights the collaborative significance of journals. Bone (1,077) and the Journal of Bone and Mineral Research (852) demonstrate the highest connectivity, indicating strong interrelations within their research networks. Other journals, like Clinical Orthopaedics and Related Research (416) and Orphanet Journal of Rare Diseases (399), show moderate influence, while journals such as Osteoarthritis and Cartilage (55) reflect minimal network integration.

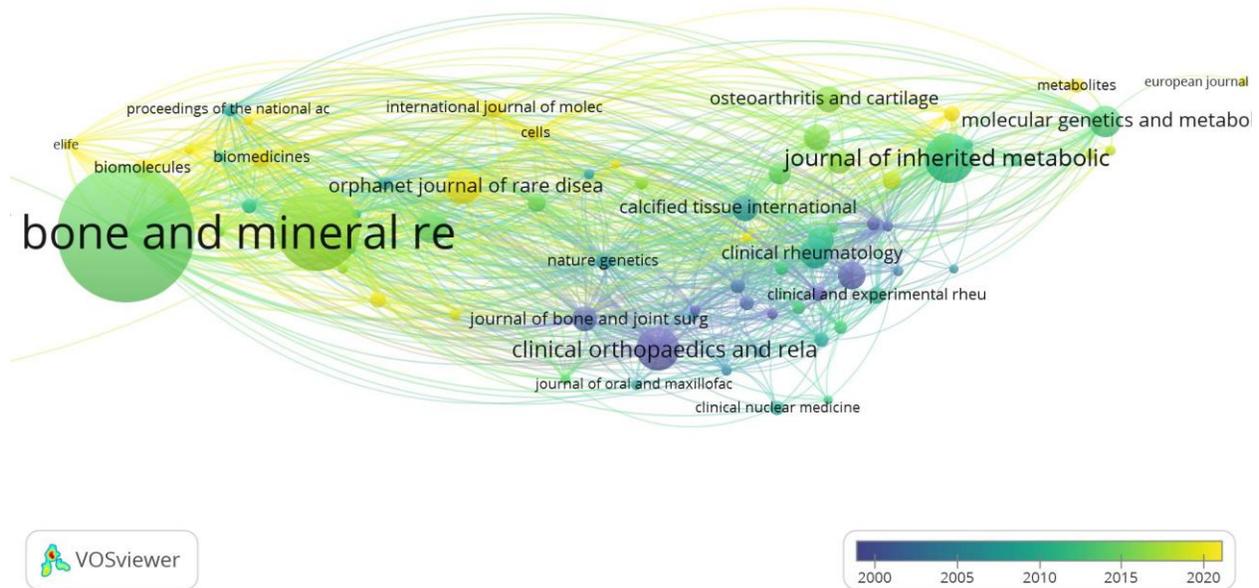

**Figure 2:** Source (journal) co-citation and collaboration network

It is worth noting that the Orphanet Journal of Rare Diseases had 26 articles published, making up 1.43% of the total contribution. However, despite a total global citation score (TGCS) of 598, there were no local citations recorded (TLCS 0). This indicates that the articles are more widely recognized on a global scale rather than being referenced locally, possibly due to its specific focus on rare diseases such as FOP. Similarly, Biomedicines and Biomolecules also lack local citations, with TGCS values of 98 and 20, respectively, suggesting that their research contributions are either emerging or highly specialized and have not yet gained widespread academic attention.

The American Journal of Medical Genetics Part A and Molecular Genetics and Metabolism, both with 22 articles each, hold significant TLCS and TGCS scores, emphasizing their relevance in the genetics and metabolic pathways of FOP. Clinical Rheumatology and Rheumatology International, with a focus on rheumatological implications, also make moderate contributions to the total with around 1% each and have notable citation impacts.

In a recent publication, The Journal of Bone and Joint Surgery-American Volume was distinguished by 16 articles (0.88%), along with notably high TLCS (843) and TGCS (2,278). These figures signify a substantial citation-per-article ratio, underscoring the journal's specialization and authority in the field of orthopedic surgery research. Conversely, Value in Health, though included, displayed low TLCS (4) and TGCS (7), indicating limited impact in the realm of FOP research.

In summary, the data shows that FOP research is heavily concentrated in a small number of influential journals, with the Journal of Bone and Mineral Research and Bone standing out for their volume and impact. However, the variation in article citations and total global citations across different journals indicates a tiered impact, where some journals act as established bases for advancing knowledge, while others are newer players with potential for more significant future influence. This distribution highlights the multi-disciplinary aspect of FOP research, encompassing genetics, orthopedics, and rare disease studies.

**Table 4: Ranking of collaborative institutions on Fibrodysplasia Ossificans Progressiva**

| S.No. | Institution | Records | % | TLCS | TGCS | ACPP |
|---|---|---|---|---|---|---|
| 1 | University of Pennsylvania | 272 | 14.97 | 4733 | 9682 | 35.60 |
| 2 | University of Liverpool | 102 | 5.61 | 1066 | 1600 | 15.69 |
| 3 | Mayo Clin | 89 | 4.90 | 373 | 751 | 8.44 |
| 4 | University of Calif San Francisco | 75 | 4.13 | 654 | 2009 | 26.79 |
| 5 | University of Siena | 69 | 3.80 | 926 | 1560 | 22.61 |
| 6 | Royal Liverpool University Hospital | 56 | 3.08 | 676 | 1053 | 18.80 |
| 7 | Ipsen | 41 | 2.26 | 52 | 81 | 1.98 |
| 8 | Royal Natl Orthopaed Hospital | 37 | 2.04 | 95 | 259 | 7.00 |
| 9 | Hospital Italiano Buenos Aires | 36 | 1.98 | 74 | 208 | 5.78 |
| 10 | Childrens Hospital Philadelphia | 35 | 1.93 | 1099 | 1935 | 55.29 |
| 11 | Saitama Medical University | 35 | 1.93 | 665 | 1761 | 50.31 |
| 12 | University of Oxford | 35 | 1.93 | 510 | 1736 | 49.60 |
| 13 | Vrije University of Amsterdam | 30 | 1.65 | 65 | 496 | 16.53 |
| 14 | Harvard University | 29 | 1.60 | 678 | 3821 | 131.76 |
| 15 | Hop Necker Enfants Malad | 28 | 1.54 | 488 | 881 | 31.46 |
| 16 | Regeneron Pharmaceut Inc | 28 | 1.54 | 81 | 775 | 27.68 |
| 17 | Slovak Academy of Science | 28 | 1.54 | 283 | 482 | 17.21 |
| 18 | University of Amsterdam | 25 | 1.38 | 66 | 274 | 10.96 |
| 19 | University of Tokyo | 25 | 1.38 | 272 | 859 | 34.36 |
| 20 | University of Genoa | 24 | 1.32 | 164 | 539 | 22.46 |

Table 4 presents the leading 20 organizations that are actively engaged in collaborating and making contributions to research on Fibrodysplasia Ossificans Progressiva (FOP). These organizations differ in terms of how much they publish, their citation impact, and academic productivity, which is gauged by the Average Citation Per Publication (ACPP), demonstrating their influence and significance in progressing FOP-related studies.

The University of Pennsylvania stands out in the field with 272 publications, making up 14.97% of all records. It has achieved notable citation metrics, such as a Total Local Citation Score (TLCS) of 4,733 and a Total Global Citation Score (TGCS) of 9,682. With an ACPP of 35.60, it emphasizes the significant impact and excellence of research produced by this institution, positioning it as a frontrunner in FOP research.

The University of Liverpool and the Mayo Clinic have made significant contributions, accounting for 5.61% and 4.90% of publications in the field. The University of Liverpool shows strong citation performance with a total local citation score (TLCS) of 1,066, total global citation score (TGCS) of 1,600, and an average citation per publication (ACPP) of 15.69. On the other hand, the Mayo Clinic has a lower citation impact with an ACPP of 8.44, indicating a more specialized or niche focus in FOP research.

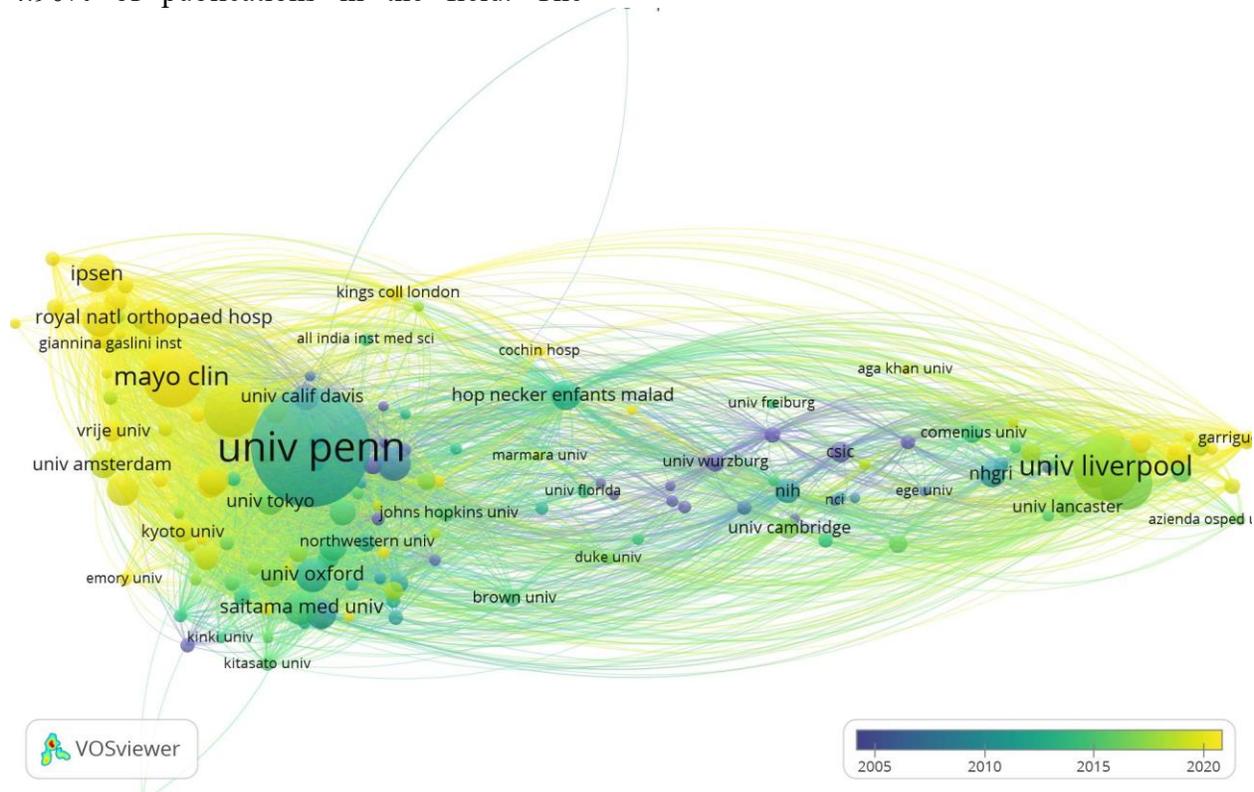

**Figure 3:** Institutional collaboration network on FOP research

The Total Link Strength (TLS) represented in Figure 3 is the node strength of institutions in the collaborative network on Fibrodysplasia Ossificans Progressiva research. The University of Pennsylvania stands out with the highest node strength of 12,342, reflecting its central role in collaborations. Similarly, the University of Liverpool (4,111) and Mayo Clinic (3,606) show substantial node strengths, indicating their significant contributions to the network. Other institutions like the University of California San Francisco (3,976) and University of Siena (3,103) also demonstrate notable connectivity, while institutions such as Ipsen (731) have comparatively lower node strengths.

Numerous institutions, such as the University of California, San Francisco (75 publications, ACPP 26.79) and the University of Siena (69 publications, ACPP 22.61), exhibit a balance between the quantity and impact of their publications, signifying their importance in terms of both research output volume and quality. Similarly, the Royal Liverpool University

Hospital has generated 56 publications (3.08%) with commendable citation metrics (ACPP 18.80), demonstrating a similar equilibrium between quantity and impact.

Some other prominent establishments worth mentioning are the Children's Hospital of Philadelphia (35 publications, ACPP 55.29) and Saitama Medical University (35 publications, ACPP 50.31). These institutions have notably high ACPP scores, suggesting that their work is extensively cited and has a substantial impact. Additionally, the University of Oxford (35 publications, ACPP 49.60) showcases its influence in the field through strong citation metrics (TLCS 510, TGCS 1,736).

It is noteworthy that prestigious establishments such as Harvard University (29 publications, ACPP 131.76) and Regeneron Pharmaceuticals Inc. (28 publications, ACPP 27.68) demonstrate remarkably high ACPP values, indicating that despite their lower publication numbers, their research has made a noteworthy impact in academia and real-world applications. Certain companies, like Ipsen (with 41 publications and an average citations per publication of 1.98), demonstrate a modest citation performance despite making a significant number of publications. This may suggest that the company is engaging in new research areas or prioritizing initial studies with less immediate influence.

Establishments from various geographical areas, like the Hospital Necker-Enfants Malades in France (TLCS 488, TGCS 881, ACPP 31.46), and the Slovak Academy of Sciences (ACPP 17.21), underscore the international scope of FOP research. Specifically, the University of Tokyo (25 publications, ACPP 34.36) and the University of Genoa (24 publications, ACPP 22.46) are noteworthy contributors from Asia and Europe, respectively.

In brief, Table 4 demonstrates the prominent influence of institutions like the University of Pennsylvania, University of Liverpool, and University of California, San Francisco in pushing forward FOP research. Institutions with elevated ACPP, such as Harvard University, Children's Hospital of Philadelphia, and Saitama Medical University, emphasize the substantial scholarly impact of their input. This distribution emphasizes the collaborative and global endeavor to progress the comprehension and management of FOP.

The distribution of research contributions on Fibrodysplasia Ossificans Progressiva (FOP) by department, as shown in Table 5, emphasizes the prominence of various academic and research institutions. The University of Pennsylvania stands out as a major player, with multiple departments making extensive contributions to FOP research. The School of Medicine tops the list with 102 records, the highest among all departments, and scores a Total Local Citation Score (TLCS) of 2,932 and a Total Global Citation Score (TGCS) of 5,914. The department's impressive Average Citations Per Publication (ACPP) of 57.98 underscores its strong scholarly impact. Other units within the university, such as the Perelman School of Medicine, the Department of Medicine, and the Department of Orthopaedic Surgery, also make significant contributions. The Center for Research on FOP & Related Disorders at the same university, while producing fewer outputs, has gained notable recognition for its focused work on this rare disease.

Harvard University's School of Medicine has a relatively low number of 18 records, but its ACPP of 174.00 is exceptionally high, indicating that the research it does is groundbreaking and highly influential. This emphasizes the importance of quality over quantity. Similarly, the University of California, San Francisco (UCSF) and the University of Siena also have strong citation metrics and balanced contributions. UCSF's Department of Medicine and its Institute for Human Genetics bring interdisciplinary expertise, while Siena's Department of Biotechnology, Chemistry, and Pharmacology consistently produces impactful research.

The Mayo Clinic, known for its exceptional clinical skills, has also been influential in the medical field with its Department of Medicine, generating 52 publications. While its ACPP of 8.02 is comparatively moderate compared to top departments, it demonstrates a commitment to practical and applied medical research. Furthermore, institutions such as the Saitama Medical University Research Center for Genomic Medicine and the Royal National Orthopaedic Hospital Center for Metabolic Bone Diseases have made substantial contributions, particularly in specialized areas of FOP research.

**Table 5: Ranking of department wise distribution on Fibrodysplasia Ossificans Progressiva**

| S.No. | Institution with Sub Division | Records | % | TLCS | TGCS | ACPP |
|---|---|---|---|---|---|---|
| 1 | University of Pennsylvania, Sch Med | 102 | 5.61 | 2932 | 5914 | 57.98 |
| 2 | University of Pennsylvania, Perelman Sch Med | 87 | 4.79 | 874 | 1683 | 19.34 |
| 3 | University of Pennsylvania, Dept Med | 57 | 3.14 | 790 | 1746 | 30.63 |
| 4 | Mayo Clin, Dept Med | 52 | 2.86 | 195 | 417 | 8.02 |
| 5 | University of Pennsylvania, Ctr Res FOP & Related Disorders | 51 | 2.81 | 633 | 1069 | 20.96 |
| 6 | University of Pennsylvania, Dept Orthopaed Surg | 50 | 2.75 | 728 | 1796 | 35.92 |
| 7 | Ipsen | 40 | 2.20 | 52 | 81 | 2.03 |
| 8 | Univ Liverpool, Inst Ageing & Chron Dis | 32 | 1.76 | 353 | 555 | 17.34 |
| 9 | Univ Calif San Francisco, Dept Med | 29 | 1.60 | 121 | 501 | 17.28 |
| 10 | University of Pennsylvania, Dept Genet | 29 | 1.60 | 652 | 1438 | 49.59 |
| 11 | Univ Calif San Francisco, Inst Human Genet | 28 | 1.54 | 113 | 428 | 15.29 |
| 12 | Univ Siena, Dipartimento Biotecnol Chim & Farm | 26 | 1.43 | 437 | 762 | 29.31 |
| 13 | Royal Natl Orthopaed Hosp, Ctr Metab Bone Dis | 24 | 1.32 | 34 | 139 | 5.79 |
| 14 | University of Pennsylvania | 23 | 1.27 | 23 | 32 | 1.39 |
| 15 | Hosp Italiano Buenos Aires, Dept Pediat | 22 | 1.21 | 61 | 146 | 6.64 |
| 16 | Saitama Med Univ, Res Ctr Genom Med | 21 | 1.16 | 253 | 482 | 22.95 |
| 17 | Univ Siena, Dept Biotechnol Chem & Pharm | 21 | 1.16 | 103 | 236 | 11.24 |
| 18 | Univ Calif San Francisco, Div Endocrinol & Metab | 19 | 1.05 | 48 | 181 | 9.53 |
| 19 | Harvard Univ, Sch Med | 18 | 0.99 | 511 | 3132 | 174.00 |
| 20 | Inst IMAGINE, Dept Genet | 18 | | 32 | 100 | |

It is intriguing to note that Ipsen, a pharmaceutical firm, brings attention to the influence of the industry in FOP research. Despite Ipsen's relatively low ACPP of 2.03, its involvement demonstrates the conversion of research into practical treatments. This emphasizes the significance of academic and industry collaboration in progressing FOP treatment options.

In the field of FOP research, the University of Pennsylvania stands out as a key player, with strong support from other top institutions including Harvard, UCSF, and Siena. Each of these institutions contributes distinct strengths to the research landscape. The significant impact of their respective departments and the participation of industry partners such as Ipsen highlight the collaborative and interdisciplinary nature of FOP research. This distribution underscores the importance of both academic rigor and translational research in addressing the complexities of this rare medical condition.

Table 6: Document type contribution of research

| S.No. | Document type | Records | % | TLCS | TGCS |
|---|---|---|---|---|---|
| 1 | Article | 1027 | 56.52 | 11108 | 30851 |
| 2 | Meeting Abstract | 265 | 14.58 | 34 | 46 |
| 3 | Review | 242 | 13.32 | 1258 | 11524 |
| 4 | Editorial Material | 131 | 7.21 | 240 | 703 |
| 5 | Letter | 76 | 4.18 | 279 | 486 |
| 6 | Article; Proceedings Paper | 28 | 1.54 | 125 | 919 |
| 7 | Note | 18 | 0.99 | 212 | 326 |
| 8 | Article; Early Access | 9 | 0.50 | 0 | 7 |
| 9 | Correction | 8 | 0.44 | 7 | 22 |
| 10 | Article; Book Chapter | 4 | 0.22 | 1 | 59 |
| 11 | Review; Book Chapter | 3 | 0.17 | 5 | 42 |
| 12 | Editorial Material; Early Access | 2 | 0.11 | 0 | 0 |
| 13 | Reprint | 2 | 0.11 | 9 | 46 |
| 14 | Discussion | 1 | 0.06 | 1 | 1 |
| 15 | News Item | 1 | 0.06 | 0 | 0 |
|  | **Total** | **1817** | **100.00** | **13279** | **45032** |

The data in Table 6 illustrate that the majority of research contributions on Fibrodysplasia Ossificans Progressiva (FOP) are in the form of research articles, making up 56.52% of the total publications (1,027 records). These articles not only represent the largest portion of research output but also have the highest Total Local Citation Score (TLCS) of 11,108 and Total Global Citation Score (TGCS) of 30,851, demonstrating their significant impact and recognition within the academic and research community. This underscores the important role of peer-reviewed articles in advancing the understanding of FOP.

The category of reviews ranks as the second most important, making up 13.32% of the records with a total of 242 documents. These reviews, which have a TLCS of 1,258 and a TGCS of 11,524, play a crucial role in consolidating current knowledge and influencing the trajectory of future research. Their high citation scores highlight their significance as sources of reference for the research community.

The portion of "Meeting abstracts" makes up 14.58% of the total records, which includes 265 documents. However, their citation metrics are relatively low with a TLCS of 34 and TGCS of 46. This indicates that while they are useful for sharing initial findings and promoting collaboration, their lasting scholarly influence is not as strong as that of full articles and reviews.

"Editorial content" and "correspondence" also play a significant role in the research content, making up 7.21% (131 documents) and 4.18% (76 documents) of the total. These types of documents frequently offer expert viewpoints, analysis, or short communications that enhance the conversation about FOP. It is worth noting that letters have a relatively high TLCS (279) in comparison to their quantity, indicating a focused impact within certain research circles.

Different types of documents, including "proceedings papers, notes, corrections, and book chapters", account for a smaller portion of the overall output but serve specific purposes. Articles that fall into both the categories of "proceedings papers" and "book chapters" tend to have moderate citation scores, suggesting their importance in certain contexts such as conferences or specialized collections. Early access articles and reprints make minimal contributions, but they play a role in disseminating emerging research and knowledge.

It is noteworthy that infrequent types like "news items, discussions, and editorial content with advanced access" comprise a small fraction of the data. Their minimal citation scores indicate a lack of significant academic influence, yet hint at their involvement in the peripheral spread of information.

In summary, the predominance of articles and reviews highlights the critical importance of traditional academic publishing in FOP research. Simultaneously, the contributions of other document types emphasize the diversity of formats utilized to share knowledge, foster collaboration, and enhance discourse in this specialized field.

The analysis of the most relevant countries by corresponding authors, presented in Table 7, highlights significant geographical trends in Fibrodysplasia Ossificans Progressiva (FOP) research. The United States leads with 678 records, representing 37.31% of the total output, and achieves the highest Total Local Citation Score (TLCS) of 7,797 and Total Global Citation Score (TGCS) of 27,273. This dominance underscores the USA's central role in FOP research, reflecting the availability of resources, infrastructure, and leading institutions driving impactful studies.

sThe United Kingdom ranks second with 318 records (17.50%), contributing significantly to the scholarly impact with a TLCS of 2,246 and TGCS of 7,161. Italy, with 161 records (8.86%), and Japan, with 126 records (6.93%), also exhibit substantial contributions, supported by notable TLCS values of 1,280 and 1,371, respectively. These countries showcase active research communities and expertise in rare diseases like FOP.

France and Germany follow closely, accounting for 6.66% (121 records) and 6.33% (115 records) of the total publications, respectively. Both countries demonstrate strong academic output, with France achieving a TLCS of 845 and Germany contributing a TGCS of 2,293, highlighting the global recognition of their research efforts.

Other European nations, such as the Netherlands, Turkey, Spain, and Slovakia,

contribute between 2% and 5% of the total records, reflecting their active participation in the field. Slovakia's TLCS of 491 and TGCS of 789 are particularly notable given its smaller overall contribution, suggesting the high quality and relevance of its research.

Table 7: Most relevant countries by corresponding authors

| S.No. | Country | Records | % | TLCS | TGCS |
|---|---|---|---|---|---|
| 1 | USA | 678 | 37.31 | 7797 | 27273 |
| 2 | UK | 318 | 17.50 | 2246 | 7161 |
| 3 | Italy | 161 | 8.86 | 1280 | 3355 |
| 4 | Japan | 126 | 6.93 | 1371 | 4341 |
| 5 | France | 121 | 6.66 | 845 | 2143 |
| 6 | Germany | 115 | 6.33 | 765 | 2293 |
| 7 | Netherlands | 97 | 5.34 | 303 | 2173 |
| 8 | Peoples R China | 84 | 4.62 | 212 | 1309 |
| 9 | Turkey | 79 | 4.35 | 323 | 612 |
| 10 | India | 71 | 3.91 | 173 | 508 |
| 11 | Canada | 70 | 3.85 | 301 | 2243 |
| 12 | Spain | 55 | 3.03 | 491 | 1885 |
| 13 | Australia | 42 | 2.31 | 860 | 2489 |
| 14 | Argentina | 41 | 2.26 | 74 | 228 |
| 15 | Slovakia | 41 | 2.26 | 491 | 789 |
| 16 | Brazil | 38 | 2.09 | 219 | 437 |
| 17 | Sweden | 28 | 1.54 | 216 | 877 |
| 18 | Belgium | 23 | 1.27 | 110 | 1101 |
| 19 | Switzerland | 20 | 1.10 | 71 | 262 |
| 20 | Poland | 17 | 0.94 | 13 | 77 |

Emerging contributions from countries like China, India, Brazil, and Argentina indicate growing interest and investment in FOP research in non-Western regions. China's 84 records (4.62%) and India's 71 records (3.91%) point to the increasing integration of these nations into the global research landscape, although their citation scores suggest room for greater international impact.

Australia and Canada, while contributing modestly in terms of records (42 and 70, respectively), display strong TGCS values (2,489 for Australia and 2,243 for Canada), indicating high-quality research with significant global influence.

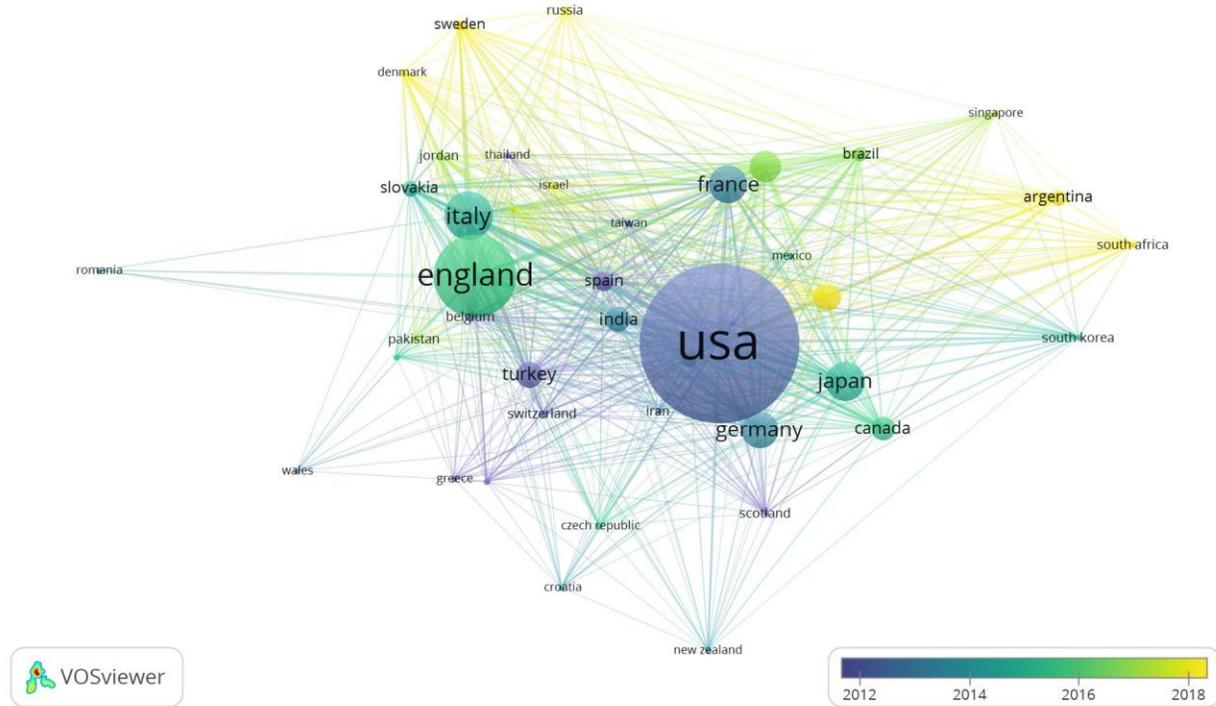

**Figure 4:** Country-level collaboration and publication output map

The Total Link Strength (TLS), as shown in figure 4, highlights the USA (13,260) and the UK (6,779) as central hubs in international research collaboration, with Italy, Japan, and France also playing significant roles. Lower TLS values for countries like Turkey (1,074) and India (799) indicate more limited global connections. Smaller contributors, such as Sweden, Belgium, Switzerland, and Poland, demonstrate niche expertise in FOP research. Sweden's TLCS of 216 and TGCS of 877, for example, highlight impactful contributions despite a lower record count (28 records, 1.54%).

In a nutshell, the data reveal a highly uneven distribution of FOP research, with the USA and European nations leading the field. However, increasing contributions from emerging research hubs in Asia and Latin America signal a positive trend toward a more globally distributed research landscape. This underscores the importance of fostering international collaboration to enhance the diversity and impact of FOP research worldwide.

**Discussion**

According to the analysis, the United States is the primary influencer in FOP research, responsible for the majority of publications (37.31%) and citation impact (TLCS of 7,797 and TGCS of 27,273). This aligns with the abundant research funding and advanced infrastructure in the US, facilitating innovative research in rare diseases such as FOP. Other significant contributors include the United Kingdom, Italy, Japan, and France, with the UK consistently holding the second position in research output and citations, reflecting substantial academic engagement in the field.

The primary focus of FOP research is found in top-tier journals within the fields of bone and mineral research, rheumatology, and genetics. The Journal of Bone and Mineral Research stands out as the leading journal in terms of both number of publications (143) and citations (TLCS of 808), indicating that advancements in bone biology, musculoskeletal disorders, and genetic research are the main drivers of FOP research.

Collaboration between institutions is essential for the advancement of FOP research. Leading institutions like the University of Pennsylvania, University of Liverpool, and Mayo Clinic have a high number of records and citation impact in this field. This demonstrates the interdisciplinary and collaborative approach of FOP research, with academic medical centers, genetic research centers, and hospitals making significant contributions.

The predominant types of documents in this field consist of "articles and reviews". Articles make up more than 50% of all publications, indicating a clear focus on conducting original research. Reviews, which account for 13.32% of the publications, play a vital role in summarizing and synthesizing the growing FOP research, providing valuable insights for researchers and clinicians. The incorporation of "meeting abstracts" underscores the importance of conferences and symposiums in disseminating initial findings and fostering collaborations.

From a geographical standpoint, the majority of FOP research is heavily focused in the United States and Europe, particularly the UK, Italy, and Germany, which have made substantial contributions. Nevertheless, emerging research from nations such as China, India, and Brazil indicate a movement towards a more worldwide research network, creating possibilities for international cooperation and the sharing of knowledge.

**Conclusion**

An analysis of research output shows that the United States is the primary driver of Fibrodysplasia Ossificans Progressiva (FOP) research, with significant contributions also coming from key European nations. This highlights the importance of collaborative efforts between academic and medical institutions specializing in genetics, musculoskeletal disorders, and rare diseases. While the USA and the UK are the main players, there is an increasing global participation in FOP research, particularly from countries in Asia and Latin America, which enriches the overall research landscape. The findings also underscore the importance of scholarly articles in furthering the comprehension of FOP, with a handful of influential journals leading the way in the field. The examination of different types of documents underscores the significance of original research and reviews in shaping upcoming paths. In general, this investigation presents a thorough summary of the current status of FOP research and offers insightful information for policymakers, researchers, and clinicians seeking to improve cooperative endeavors and progress scientific knowledge in this uncommon ailment.

Based on the findings of this bibliometric analysis, it is recommended that:

1. Future FOP research strengthens international and interdisciplinary collaborations to enhance knowledge exchange and accelerate scientific progress. Research institutions with lower output and citation impact should seek partnerships with highly productive centers such as the University of Pennsylvania and the University of Liverpool.
2. There is also a need to expand research in underrepresented regions, particularly in Asia, Africa, and South America, to improve global representation. Journals with emerging contributions should aim to increase visibility through collaborative special issues on rare diseases.

3. Finally, more clinical, genetic, and translational research is needed to address gaps in diagnosis, management, and treatment strategies for FOP.

**Conflict of interest**

There is no conflict of interest reported by the authors.